\documentclass[onefignum,onetabnum]{siamart171218}

\usepackage[english]{babel}
\usepackage{amsmath}
\usepackage{amsfonts}
\usepackage{siunitx}
\usepackage{algorithmic}
\usepackage{xcolor}
\usepackage{graphicx}
\usepackage{setspace}
\usepackage[normalem]{ulem}

\newcommand{\profresult}[2]{#2}

 \newcommand\EdRem[1]{}

\headers{
Optimizing AIREBO
}{
Markus H\"ohnerbach, and Paolo Bientinesi
}

\title{
Optimizing AIREBO: Navigating the Journey from Complex Legacy Code to High Performance
\thanks{
Submitted to the editors DATE. 
\funding{
The authors gratefully acknowledge financial support from the Deutsche Forschungsgemeinschaft (German Research Association) through grant GSC 111, and from Intel via the Intel Parallel Computing Center initiative.
}}}

\author{
Markus H\"ohnerbach
\thanks{
RWTH Aachen University, Germany
  (\email{hoehnerbach@aices.rwth-aachen.de}, \email{pauldj@aices.rwth-aachen.de}).}
\and
Paolo Bientinesi
\footnotemark[2]}

\ifpdf
\hypersetup{
  pdftitle={Optimizing AIREBO: Navigating the Journey from Complex Legacy Code to High Performance},
  pdfauthor={Markus Hoehnerbach, and Paolo Bientinesi}
}
\fi

\begin{document}
\maketitle

\begin{abstract}
Despite initiatives to improve the quality of scientific codes,
there still is a large presence of legacy code.
Such code often needs to implement a lot of functionality under time constrains, sacrificing quality.
Additionally, quality is rarely improved by optimizations for new architectures.
This development model leads to code that is increasingly difficult to work with.
Our suggested solution includes complexity-reducing refactoring and hardware abstraction.

We focus on the AIREBO potential from LAMMPS, where the challenge is that any potential kernel is rather large and complex, hindering systematic optimization.
This issue is common to codes that model multiple physical phenomena.

We present our journey from the C++ port of a previous Fortran code to
performance-portable, KNC-hybrid, vectorized, scalable, optimized code supporting full and reduced precision.
The journey includes extensive testing that fixed bugs in the original code.
Large-scale, full-precision runs sustain speedups of more than 4x (KNL) and 3x (Skylake).
\end{abstract}



\section{Introduction}

Despite the current initiatives to improve the software quality of scientific codes,
the field of scientific computing is still vastly populated by legacy code.
Such code is often
developed by domain scientists who are likely to focus on functionality rather than software quality.
The situation is worsened by the constant need not only to maintain but also to update the code to exploit the parallelism and the architectural features offered by every new generation of hardware. 
In practice, the current software  development model delivers code that is increasingly difficult to maintain, adapt, and extend.
In our opinion, the solution is to be found in the adoption of modern software engineering best practices and in the abstraction from the specific target hardware.
The case that we are presenting here, the AIREBO potential \cite{stuart_reactive_2000}, included in the
LAMMPS molecular dynamics simulation program \cite{plimpton_fast_1995},
suffers from exactly that complexity.
This paper presents our experience of refactoring the code to a point where
optimization becomes possible, while ensuring correctness of the result, and
achieving sizable speedups on state-of-the-art systems.

LAMMPS is a code widely used by computational chemists, biologists and
materials scientists to run molecular dynamics (MD) simulations \cite{rapaport_art_2004}.
MD is a popular method to bridge the gap between
continuum simulation such as the finite element method (FEM) and quantum methods such as density
function theory (DFT). In a
nutshell, it is a method to simulate systems of atoms and their movement
timestep by timestep.
As simulations can reach the scale of millions of atoms and billions of timesteps,
MD is responsible for a sizeable amount of the compute time spent on
supercomputers.

LAMMPS' primary mode of parallelization is an MPI-based domain decomposition.
This allows scaling from individual cores to entire supercomputers with little loss of efficiency.
Consequently, our efforts focus not on distributed memory parallelism, but on the lower levels:
shared memory parallelism, accelerator support, and especially vectorization.

The AIREBO potential provides an empirical model of the inter-molecular and intra-molecular forces of hydrocarbons.

As such, it is frequently used when simpler
 MD force fields
such as CHARMM, AMBER or GROMOS fail, but the input systems are too large to use 
methods such as DFT~\cite{fiolhais_primer_2003}.
Indeed, while DFT is limited to hundreds of atoms due to its super-linear
scaling behaviour, the linear scalability of MD simulations allows for much larger systems.
The aforementioned force fields
all require a user-provided system topology that prescribes the location and behaviour of bonds, angles and dihedrals.

The only non-prescribed potentials that commonly act between pairs of atoms and decay
with distance.
Many phenomena, including gravity, electrostatics and van-der-Waals forces,
can be described with such potentials.

However, some phenomena, such as the interactions in metals or of bonding
behaviour, depend on the surrounding arrangement of atoms rather than just the distance.
These are so-called many-body potentials, where interactions freely occur not just
between pairs of atoms, but also triples of atoms, quadruples of atoms, and so
on.
Many-body potentials are a lot more costly on an atom-by-atom basis than pair-potentials, and pose unique challenges due to their nature:
They contain many more terms, deep loop-nests to compute properties based on e.g., angles, and the loops tend to be rather short.
The computation itself tends to be much more expensive:
Where van-der-Waals forces can be calculated with a about two dozen FLOPS, any many-body potential requires hundreds if not thousands of FLOPS.

AIREBO is such a many-body potential. It's only input are the atom positions, and the topology is deduced from local neighborhoods.
Thus, the system topology is dynamic, rather than static, and it is possible to simulate the breaking and
formation of bonds.
Although powerful, this feature comes at a significant cost:
LAMMPS' AIREBO implementation is organized into kernels,
as typically advocated in the HPC community; however, the complexity of said kernels is such that systematic work within the code is hindered.
This situation is by no means isolated to AIREBO; in fact, 
it occurs in all cases where no one single portion of the code is responsible
for the vast majority of the execution time \cite{coon_managing_2016} \cite{kylasa_reactive_nodate}.
In AIREBO, the spread of compute time over various routines comes from the many different physical phenomena handled by the code, be it dihedrals, bond-orders, radicals or dispersion.
And while in most cases the inter-molecular force calculation dominates the computation time, 
once speed-ups are attained for this routine, the bottleneck shifts to other routines.

To conduct systematic optimization within this challenging environment, we used the following approach.
From our starting point, which was the C++ port of a Fortran implementation, we gradually isolated the code, tested and refactored it;
the testing efforts uncovered a number of issues in the existing code, which
were fixed in the course of our work.
The refactoring made it possible to work on optimization and vectorization.
At first, this was done for one single architecture, to allow for the lowest overheads, simplest debugging and highest performance.
In a separate step, we used source-to-source transformation and a vector
abstraction library to make the code performance-portable for x86 generations with AVX,
AVX2, IMCI and AVX-512 support, 
effectively covering all x86 machines currently in service at supercomputing centers across the world.
In addition to being optimized and vectorized, our code also supports
reduced precision modes and offloading to Xeon Phi accelerators.

In addition to the challenges presented by the complex nature of the code,
there also exist challenges from the perspective of vectorization, 
which include 1) data-dependent, unlikely branches, 2) deep loop nests with
low trip counts (much lower than the vector length), and in particular, 3) searches through the inferred molecular structure.
Our final, vectorized code achieves speedups of up to 8x on Intel's KNL platform, and 6x on Skylake at reduced precision.
In double precision, large-scale runs on up to 8700 cores yield sustained speedups of more than 4x (KNL) and 3x (Skylake).
  
\paragraph{Contributions}
\begin{itemize}
\item A step by step description of our approach to testing and debugging a
  simulation code that operates on millions of particles in complex manners (Sec.~\ref{ssec:test}).
\item
  An explanation of how code refactoring leads to opportunities for
  optimizations.{
  A discussion of code refactoring, code duplication, and optimization opportunities (Sec.~\ref{sec:refntest}).}
\item
  An approach for performance portable optimizations based on source-to-source
  transformation (Sec.~\ref{ssec:port}, \S 2).

\item A number of techniques 
  to significantly optimize the code either directly or by enabling vectorization (Sec.~\ref{sec:opt}).
\item Some feature suggestions for{A proposal for both} OpenMP SIMD or{features and} a unified vector abstraction library (Sec.~\ref{ssec:port}, \S 4).
\end{itemize}

\paragraph{Related Work}

Besides AIREBO, there are a number of other ``many-body'' potentials---e.g., Tersoff
\cite{tersoff_new_1988}, Stillinger-Weber \cite{stillinger_computer_1985}, EAM
\cite{daw_embedded-atom_1984}, COMB \cite{liang_classical_2013}, SNAP
\cite{thompson_spectral_2015} and ReaxFF \cite{chenoweth_reaxff_2008}. 
Without exception, their vectorization will pose
issues similar to those presented here.

Besides LAMMPS, many molecular dynamics packages aside have been optimized for new hardware.
These include Gromacs \cite{kutzner_best_2015}, NAMD \cite{stone_gpu-accelerated_2010}, DL POLY, and ls1 mardyn \cite{heinecke_performance_2015}.
Gromacs in particular has a highly portable scheme for the vectorization of
pair potentials that leads to a best-in-class single-threaded performance \cite{pall_flexible_2013}.
LAMMPS had a few efforts to incorporate new programming models, such as support for GPUs, OpenMP, vectorization/accelerators, and KOKKOS.
One of the most comprehensive studies of vectorization for MD simulations considered miniMD, a proxy for the pair-wise forces in LAMMPS \cite{pennycook_evaluating_2012} \cite{pennycook_exploring_2013}.
However, most efforts focused on GPUs, which share some, but not all of the
issues of the optimization for vector units \cite{anderson_general_2008}.
In particular, vectorized implementation of multi-body potentials are rare \cite{brown_optimizing_2015} \cite{hohnerbach_vectorization_2016}, whereas GPU implementations are more common \cite{fan_efficient_2017} \cite{nguyen_gpu-accelerated_2017} \cite{kylasa_reactive_nodate} \cite{brown_implementing_2013}.
There has been considerable work to optimize multi-body potentials in general, for example for the ReaxFF potential \cite{aktulga_parallel_2012}.

In recent years, the HPC community in general, and the MD simulation community in particular, have become more and more aware of the issues of both software quality and performance portability \cite{lucas_top_2014} \cite{bartlett_xsdk:_2017} \cite{bartlett_xsdk_2017} \cite{dubey_design_2015}.
Often, such initiatives come paired with considerations regarding optimization
for future exascale computing systems \cite{markidis_tackling_2015}.
A systematic verification---analytically using Mathematica---of the REBO
implementation in LAMMPS was carried out \cite{favata_analytical_2016}.

\paragraph{Organization of the paper}

Sec.~\ref{sec:background} provides a short overview of MD simulations in
general, and AIREBO in particular.
Sec.~\ref{sec:refntest} is a recount of our refactoring and testing approach. 
Building on that, various optimizations are discussed in Sec.~\ref{sec:opt}.
A performance evaluation across a
range of x86 generations, both at the node level and at scale, is presented in
Sec.~\ref{sec:results}. Finally, conclusions are drawn in Sec.~\ref{sec:conclusion}.

\section{Molecular Dynamics}\label{sec:background}

In the most basic terms, an MD simulation
takes a system of atoms as input and
repeatedly, in discrete time-steps,

calculates the forces that act on each atom, and updates the positions and velocities of the atoms accordingly.
The force and velocity updates are dictated by an integration routine for which
many choices exist.
In all cases, the cost of these routines is negligible when
compared to that of the force calculation, which

not only represents the bottleneck in any simulation, it is also 
considerably more complex than the integration, as it consists of multiple
kernels each calculating rather involved formulas.
Because of this, the force calculation is the main target of any optimization effort, be it through algorithmic improvements, vectorization, accelerator-use, and/or parallelization.

The forces that act on the atoms are calculated according to a potential,
a scalar quantity that describes the energy inherent in the relative positions of the atoms---the potential energy.
Many different kinds of potentials exist, each appropriate for a certain kind of material.
Typically, a potential is constructed from a set of mathematical expressions
for various energy terms based on intuition and physics, whose parameters are determined by fitting against various expected properties of the material.
The forces that act on the individual atom are then nothing more than the
negative derivative of the potential with respect to that atoms position:
$\mathbf{F}_i = -\nabla_{\mathbf{x}_i} V$.
This is a famous consequence of classical mechanics, and intuitively means
that the atoms move towards a lower potential energy.

A frequent pattern that occurs when evaluating the forces due to potentials
such as AIREBO is the computation of
derivatives of a function that takes as an argument a sum over values based on
atom positions:
$A = f\left(\sum_i g(\mathbf{x}_i)\right)$.
To evaluate the forces as $\forall i \ \mathbf{F}_i = -\nabla_{\mathbf{x}_i} A$, the list of atoms has to be traversed twice:
First, the ``forward'' pass, to compute the inner sum---from which both $A$
and $A'$ (the derivative of $f$) will be derived; then, the ``reverse'' pass,
to use $A'$ to compute all the forces by means of the chain rule:
$\mathbf{F}_i = -A'\cdot\nabla_{\mathbf{x}_i} g(\mathbf{x}_i).$
As a side note, the nature of this calculation is similar to automatic
differentiation without checkpointing/{ or }a tape \cite{bartholomew-biggs_automatic_2000}.

To speed up the computation of the forces,
a common technique is to construct neighbor lists:
One prescribes a cutoff distance beyond which the strength of the interaction is negligible,
and for each atom collects 
a list containing all other atoms within that distance.
The potential is then calculated by using only the
neighbor list elements.

Neighbor lists can be constructed in two ways:
Either by calculating the distance between all pairs of
atoms and comparing it against the cutoff (quadratic time and constant space complexity), or by spatially binning the atoms and then only considering adjacent bins (linear time and space complexity).
A neighbor list with a certain cutoff can also be used to construct
other neighbor lists with smaller cutoffs.
Besides distance, neighbor lists might include or exclude atoms according to
other criteria, e.g., if the neighbor list of atom $i$
contains atom $j$, then the neighbor list of atom $j$ should not contain atom
$i$, so that the simulation only computes each interaction once.

If the neighbor list had to be constructed at each timestep, there would be
little gain if at all.
However, 
since the distance an atom can travel in a single timestep is bounded,
it is possible to add a small ``skin'' distance to the cutoff, and
only reconstruct the neighbor list if some atom has
moved farther than that. 
Effectively, this optimization makes it possible to only reconstruct the
neighbor list once every $\approx$10 timesteps.

The longer-ranged neighbor list in AIREBO (Sec.~\ref{sssec:longer}) is built (with skin) by binning, the short-ranged list is built from that one (without skin).

\subsection{AIREBO}

The AIREBO potential was developed to enable accurate simulations of materials consisting of Hydrogen and Carbon.
In particular, this potential enables the modelling of the bonding
behaviour 
without relying on a prescribed system topology, inferring the bonds
based on proximity and relative position of the atoms.
The key extension over its precursor, the REBO potential \cite{brenner_second-generation_2002}, is that AIREBO
models not only the ``intra-molecular'' bonding forces that act within a
molecule, but also the ``inter-molecular'' forces between molecules.

These forces act at different 
    length-scales, 
    below \SI{2.0}{\angstrom} and
    below \SI{10.2}{\angstrom}, respectively.

  Since the volume increases with the cube of the distance, 
    one expects about 100 times many more
    interactions between molecules than within molecules.
  Because of this difference, 
  the code requires two separate neighbor lists, one for each length-scale.

Whenever we show percentages, these are relative to the time spent in the force calculation part of one benchmark simulation (polyethylene) and the original AIREBO code in LAMMPS on a single-threaded KNL machine in double precision.

As such, these numbers should be understood as illustrations of the relative weight of individual terms.

Only less than 1\
The exact fraction depends on the chosen skin distance and can be tuned by the user.
The small neighbor list needs to be constructed every timestep and
takes up \profresult{(o.tt('rebo_neigh')) / (o.tt('torsion') +
  o.tt('frebo') + o.tt('flj') + o.tt('rebo_neigh'))}{15.2}\
force calculation.

We also apply our improvements to two close relatives of AIREBO:
The aforementioned REBO potential (no inter-molecular forces),
and the AIREBO-m potential \cite{oconnor_airebo-m:_2015}, with modified inter-molecular forces.

In the following sections, we describe the individual terms of the
short-ranged (``intra-molecular'') and longer-ranged (``inter-molecular'') contributions.
The discussion is limited to the potential energy, since the force can be derived by simply taking the derivatives.

\subsubsection{Short-Ranged Contributions}

\begin{algorithm}[t]\caption{REBO force calculation. Not shown: Bond-order calculation.}\label{alg:rebo}
\begin{algorithmic}
\FOR{$i \in$ local atoms}
\FOR{$j \in$ short-ranged neighbors of $i$, $i < j$}
\STATE $b_{ij} \gets $ compute $b_{ij}$\;
\STATE $E \gets E + f_R(r_{ij}) + b_{ij} f_A(r_{ij})$\;
\STATE $f' \gets f_R'(r_{ij}) + b_{ij}f_A'(r_{ij})$\;
\STATE $\mathbf{F}' \gets f' \cdot (\mathbf{x}_i - \mathbf{x}_j) / r_{ij}$\;
\STATE $\mathbf{F}_i \gets \mathbf{F}_i - \mathbf{F}'$\;
\STATE $\mathbf{F}_j \gets \mathbf{F}_j + \mathbf{F}'$\;
\ENDFOR
\ENDFOR
\end{algorithmic}
\end{algorithm}

\begin{algorithm}[t]\caption{Torsion force calculation.}\label{alg:torsion}
\begin{algorithmic}
\FOR{$i \in$ local atoms}
\FOR{$j \in$ short-ranged neighbors of $i$, $i < j$}
\FOR{$k \in$ short-ranged neighbors of $i$, $k\neq j$}
\FOR{$l \in$ short-ranged neighbors of $j$, $l\neq k$, $l\neq i$}
\STATE $E \gets E + w_{ki}(r_{ki}) w_{ij}(r_{ij}) w_{jl}(r_{jl}) V^\textrm{tors}_{kijl}(\omega_{kijl})$\;
\STATE \dots
\STATE $\mathbf{F}_k \gets \mathbf{F}_k + \mathbf{F}'_k$\;
\STATE $\mathbf{F}_i \gets \mathbf{F}_i + \mathbf{F}'_i$\;
\STATE $\mathbf{F}_j \gets \mathbf{F}_j + \mathbf{F}'_j$\;
\STATE $\mathbf{F}_l \gets \mathbf{F}_l + \mathbf{F}'_l$\;
\ENDFOR
\ENDFOR
\ENDFOR
\ENDFOR
\end{algorithmic}
\end{algorithm}

The AIREBO potential includes two short-ranged contributions: the REBO and the
torsion terms.
Together, 
they model the bonding behaviour of the atoms.

The REBO term consists of a repulsive pair-wise term $f_R$,
and an attractive pair-wise term $f_A$ combined with the bond-order $b_{ij}$:
$E^\textrm{REBO}_{ij} = f_R(r_{ij}) + b_{ij} f_A(r_{ij})$.
Alg.~\ref{alg:rebo} illustrates how the forces are computed from this expression.
For the bond-order calculation, see Sec.~\ref{sec:bond-order}.
The torsion term
acts between four atoms, where first and second, second and third, and third and forth are neighbors (abridged implementation in Alg.~\ref{alg:torsion}).

About \profresult{o.tt('frebo') / (o.tt('frebo') + o.tt('torsion') + o.tt('flj') + o.tt('rebo_neigh'))}{3.3}\

\subsubsection{Longer-Ranged Contributions}\label{sssec:longer}

\begin{algorithm}[t]\caption{Lennard-Jones calculation. Not shown: Bondorder and
    $C_{ij}$ calculation.}\label{alg:lj}
\begin{algorithmic}
\FOR{$i \in$ local atoms}
\FOR{$j \in$ longer-ranged neighbors of $i$, $i < j$}
\STATE $C_{ij} \gets$ compute by iterating through neighbors of $i$ and their neighbors\;
\IF{$C_{ij} = 0$}
\STATE continue\;
\ENDIF
\STATE $P\gets S_r(r_{ij})$\;
\STATE $Q\gets 1$\;
\IF{$P \neq 0$}
\STATE  $B\gets $ compute $S_b(b^\textrm{*}_{ij})$\;
\STATE  $Q\gets P\cdot B + 1 - P$\;
\ENDIF
\STATE $E \gets E + Q C_{ij}V^\textrm{LJ}_{ij}(r_{ij})$\;
\STATE \dots
\STATE $\mathbf{F}_i \gets \mathbf{F}_i + \mathbf{F}'_i$\;
\STATE $\mathbf{F}_j \gets \mathbf{F}_j + \mathbf{F}'_j$\;
\ENDFOR
\ENDFOR
\end{algorithmic}
\end{algorithm}

Since it models inter-molecular forces, this contribution needs to act at a much longer range.
Because it is highly repulsive at short distances, this
contribution also needs to be switched off if significant bonding interaction exists (first factor), and if both atoms are closeby parts of the same molecule (second factor).
The actual inter-molecular potential (here $V_\textrm{LJ}$) follows either the
Lennard-Jones formula (in the AIREBO case) \cite{jones_determination_1924}, or the Morse formula (in the AIREBO-m case) \cite{morse_diatomic_1929}.

\[E^\textrm{LJ}_{ij} = (S_r(r_{ij})S_b(b^\textrm{*}_{ij}) + 1 - S_r(r_{ij})) C_{ij}V^\textrm{LJ}_{ij}(r_{ij}).\]

The first switch-off factor is composed of the two cutoff functions $S_r$ and $S_b$.
$S_r$ limits the effect of the $S_b$ function to within a certain cutoff radius.
$S_b$ is then a factor that switches off the longer-ranged interaction if $b^\textrm{*}_{ij}$ exceeds a certain value.

The second switch-off factor is denoted by $C_{ij}$:
\begin{multline*}
C_{ij} = 1 - \max_{k,l}\{w_{ij}(r_{ij}), w_{ik}(r_{ik})w_{kj}(r_{kj}),\\w_{ik}(r_{ik})w_{kl}(r_{kl})w_{lj}(r_{lj})\}.
\end{multline*}
It is computed by searching thorugh the short-ranged neighbor list for either a direct connection, a connection via a single neighbor, and via a neighbor and its neighbor.

About \profresult{o.tt('flj') / (o.tt('frebo') + o.tt('torsion') + o.tt('flj') + o.tt('rebo_neigh'))}{80.1}\
Of this, \profresult{(o.tt('flj path_forward') + o.tt('flj path_reverse')) / o.tt('flj')}{47.8}\
In \profresult{o.n('flj per_j cijeq0=1') / o.n('flj per_j cijeq0=*')}{3.2}\
In \profresult{o.n('flj per_j need_bondorder=1') / o.n('flj per_j need_bondorder=*')}{2.2}\

\subsubsection{Bond-Order Calculation}\label{sec:bond-order}

The bond-order calculation is required for the REBO part of the potential and the modified bond-order calculation in the longer-ranged contribution.
The bond-order is a measure of the strength of a bond between two atoms $i$ and $j$.
The modified bond-order differs from the unmodified one in two ways:
First, it is evaluated at a fictitious distance between atoms $i$ and $j$, i.e., every occurrence of $r_{ij}$ is replaced by a constant.
Second, it is plugged into $S_b$, a cutoff function.
The consequence is that the derivatives of the overall term are zero,
since the cutoff function is constant 0 or 1 everywhere except for its transition region.

The bond-order $b_{ij}$ consists of three different terms,
$p^{\sigma\pi}$,  $\pi^\mathrm{rc}$, and $\pi^\mathrm{dh}$,
which model covalent bonding, radical interactions, and rotations around multiple bonds:
\[b_{ij} = \tfrac{1}{2}(p^{\sigma\pi}_{ij} p^{\sigma\pi}_{ji}) + \pi^\mathrm{rc}_{ij} + \pi^\mathrm{dh}_{ij}.\]
The term $\pi^\mathrm{rc}$ is the simplest one:
It is a spline in $N^C_{ij}$ and $N^\mathrm{conj}_{ij}$, which in turn are sums over short-ranged neighbors.
The term $\pi^\mathrm{dh}_{ij}$ is a product of another such spline $T$ and a
sum over dihedrals formed by the neighbors of both atoms $i$ and $j$:
\[\pi^\mathrm{dh}_{ij} = T_{ij} \sum_k \sum_l \dots\]
Finally, the term $p^{\sigma\pi}$ consists of a sum over neighboring atoms and is dependent on their angles and a spline $P_{ij}$.
\[p^{\sigma\pi}_{ij} = [1 + \sum_k w_{ik}(r_{ik})g_i(\cos\theta_{jik})e^{\lambda_{jik}} + P_{ij}]^{-1/2}.\]

In total, \profresult{o.tt('bondorder') / (o.tt('flj') + o.tt('frebo') +
  o.tt('torsion') + o.tt('rebo_neigh'))}{8.2}\
is spent with the calculation of bond-orders.

They make up \profresult{o.tt('frebo bondorder') / o.tt('frebo')}{93.9}\
For the modified bond-order, \profresult{o.n('flj bondorder dstbneq0=1') / o.n('flj bondorder dstbneq0=*')}{0.0}\

\section{The Original Implementation}\label{sec:refntest}

The LAMMPS version of AIREBO was added in 2007;
for the remainder of this article, we refer to that implementation as the ``original'' code.
Our goal was to create alternative implementation optimized for current computer architectures.
Since the original code contains about 3000 lines of code that are involved in the actual calculation, and about 2000 lines reading parameter files and initializing various data structures, our initial goal was to isolate the computational from the administrative code, and then refactor it to enable optimization.

The biggest obstacle we faced when we set out to
  optimize the various routines was code duplication: Not in one
  case was a loop over atoms reused, despite ample opportunities.
These opportunities arise from the fact that certain tasks need to be performed multiple times on different data.
Indeed, when comparing different routines such as the bond-order and the
  modified bond-order calculations, one can identify a lot of similar code; 
the same is true also within each
  routine: For instance, the computation of different terms for $p_{ij}$ and
  $p_{ji}$ share code, as do the force updates from splines.

We now present our isolation, refactoring and testing work.

\subsection{Isolation}

Code isolation was one of our first steps towards code refactoring.
This brought many advantages:
1) it enabled us to perform testing without the overhead of calling into the simulation code;
2) it simplified automated testing, since our test-driver only needs to
interact with the code itself rather than operating on simulation output
files; 
3) it allowed us to target the specific instance in which we suspect a bug, e.g., when the bug only occurs on atom $X$ at timestep $Y$;
4) it made debugging easier since it eliminates any error-source outside of the force calculation code.
Furthermore, a high degree of isolation
is necessary for offloading to 1st gen. Xeon Phi.

  In order to achieve code isolation, we extracted
  all the input parameters and data (size, positions, neighbor lists, ...),
  the intermediate data (small neighbor list and per-atom quantities),
  and the output data (forces) 
  after the execution of each of the following steps: small neighbor list, REBO interaction, torsion interaction, and Lennard-Jones interaction.
  We then copied the code, modified it to use the captured data and made sure it still works correctly.

\subsection{Refactoring}

We first wrote a refactored code that is based on the original AIREBO implementation present in
LAMMPS, and used this as the ``reference'' code for our optimization efforts.
In particular, it was used to debug both LAMMPS' AIREBO implementation, and
our optimized implementation;
because of the way the code is decomposed into easy to reuse
functions, the task of debugging is simplified. 
This code is also used in the optimized implementation, to handle remainder loops and cases where assumptions are violated.
This design delivers best performance in the most common case, and remains
correct even in odd cases. 
In total, the refactored code has about 1500 lines of force calculation code---a 50\
Some additional initialization code  (about 700 lines) is mostly related to
offloading, and is necessary for the 1st generation Xeon Phi.

\subsection{Testing}\label{ssec:test}

During our efforts to reduce code duplication, we became wary of
the correctness of the original LAMMPS code:
On the one
  hand, in multiple occasions we found an inconsistency with respect to the reference paper \cite{stuart_reactive_2000};
  on
  the other hand, we were informed of users reporting issues with seemingly
  unidentifiable root causes.
For these reasons, we decided to use not just our newly implemented code as a base of comparison, but also looked around for other implementations.
We found such an implementation in the OpenKIM repository \cite{tadmor_potential_2011}.

Since LAMMPS supports the KIM API, it could readily be integrated (after some minor fixes).
It then became apparent that the OpenKIM-based code, from the original authors of
AIREBO, was at least in some capacity the predecessor of the LAMMPS code, and 
both shared many variable names and idiosyncrasies.
Nevertheless, the OpenKIM code proved correct in the vast majority of cases:
While we identified numerous issues in the original LAMMPS code we found just one in OpenKIM.

At that point, we were operating on three separate implementations of
  AIREBO: 
The original LAMMPS code, the OpenKIM code and our own code.
Testing was performed on a range of simulations,  from very simple model systems to production-like systems, and random systems.
Since AIREBO implementations are filled with branches and special cases, simple strategies such as small model problems and random systems often fail to uncover notable issues.
On the other hand, real systems, in particular when set up such that the
reactive regime is exercised, were most effective at exposing bugs, which
tended to manifest themselves as large discrepancies in the forces.
Thanks to this, we were able to collect traces of simulations (i.e., atom positions) from one
code, and test against the 
other two codes.

By painstakingly isolating the different contributions to the forces, we
  were able to pinpoint differences in the various codes. These differences
  motivated a manual inspection of the codes, which, when compared with the
  analytical formulas derived from the energy expressions, eventually led us to a number of
  bugs in the LAMMPS implementation and one bug in OpenKIM.

As all three implementations now agree, the confidence level in our
implementation is high.

Given the size and complexity of the original LAMMPS code, it is not surprising that
the bugs we found (and fixed) were diverse in nature and cause.
Examples from the LAMMPS code include:
the Kronecker-delta factors based on atom type were ignored, variables were
used without initialization, some energy terms were also computed while doing
the reverse iteration for derivatives, the way the $b^*_{ij}$ term is supposed
to be constructed w.r.t. $r_{ij}$ (see section~\ref{sec:bond-order}) was
ignored, in various places signs were flipped, variables were reused
incorrectly, or with different conventions for their values.

\section{Improvements}\label{sec:opt}

As mentioned previously, we aim to integrate our code in the popular, open-source LAMMPS molecular dynamics package.
LAMMPS is designed to be easily extensible: One mechanism to provide
extensions are the so-called packages, which 
allow users to add features, such as the support for
many-body potentials, at compile-time.
Performance enhancements, such as support for
OpenMP, GPU, and Intel's hardware (offloading to Xeon Phi
accelerators, and vectorization support), are also organized into packages.
In particular, our code
is part of the optimizations specific to Intel hardware.
We later comment on how these optimizations can also benefit other architectures.

  To identify  where the highest potential for speedup lies,
  code optimization should always follow a phase 
  in which the target code is profiled with various workloads.
These workloads should not just differ in size, but also in structure and composition.
 In this work,
  we used a mix of profiling tools, ranging from manually adding timers, to automated source-code annotation, to Linux' perf tool, and Intel's VTune and Amplifier tools.

When aiming for vectorization, our experience is that it pays off to leave {serial }all
the methods the code calls until{unless} proven a bottleneck. This strategy
also aids debugging, since the serial code is already well tested. 
While we tried various approaches to white-box testing the vectorized code
against known-good code based on runtime intermediate values,  a good
solution remains elusive.
However, we found it helpful to dump the simulation state across a simulation
and then rerun that state, i.e., compute forces at all the recorded sets of
atom positions, but using the dump, instead of the computed forces,  to move the atoms.
In our opinion, this is the only reliable approach for side-by-side comparisons, since otherwise the inputs will deviate in the course of the simulation.
To this end, it paid off to run not only simulations that are indicative of
production runs, but also non-equilibrated systems. Indeed, the latter
  stressed the reactive parts of the potential implementation, and turned out to be instrumental in spotting 
bugs.

As in the previous section, percentages are relative to the time spent in the
force calculation part of one benchmark simulation (polyethylene) and the original LAMMPS AIREBO code
on a single-threaded KNL machine in double precision.
As such, these numbers should be understood as illustrations of the relative weight of individual terms.
We distinguish between the speedup from vectorization and optimization.
The latter includes all the optimizations that are described, but used with a vector length of one;
it also includes all vector overheads (despite the scalar execution), and as
such is just a simple indicator of whether or not an optimization is also potentially beneficial without vectorization.
Optimization speedup is $T{\textrm{Time}}(\textrm{original LAMMPS code}) / T{\textrm{Time}}(\textrm{our scalar code})$, vectorization speedup $T{\textrm{Time}}(\textrm{our scalar code}) / T{\textrm{Time}}(\textrm{our vectorized code})$.

\subsection{Short-Ranged Neighbor List}

The calculation of the short-ranged neighbor list occurs each timestep, and 
is created from the longer-ranged neighbor list.
For the correctness of the overall code, this list is expected to be exact:
Any atom in the list needs to satisfy the cutoff.

In the original LAMMPS code, the portion
  responsible for this task takes about \profresult{o.tt('rebo_neigh') / (o.tt('flj') + o.tt('frebo') + o.tt('rebo_neigh') + o.tt('torsion'))}{15.2}\
  
That code is using the longer-ranged neighbor list to construct the short-ranged list.
Each timestep, it iterates through each entry in that list, and filters out those entries that also meet the tighter cutoff for the short-ranged list.

As a first optimization, we introduce an intermediate neighbor list by extending the short-ranged neighbor list cutoff with a skin distance.
During those timesteps where the longer-ranged neighbor list gets rebuilt, we
need to also rebuild the intermediate one;
however, in all other timesteps, to build the short-ranged neighbor list 
we only need to iterate through the intermediate one.
This optimization yields a speedup of \profresult{o.tt('rebo_neigh') / sd.tt('rebo_neigh') / 100}{3.3}x.

In addition, we vectorize the calculation of the neighbor list.
This can be achieved by either building multiple neighbor lists
simultaneously,
or by inserting at once multiple elements into a single neighbor list.
Benchmarking revealed that the latter option performs better, giving an
additional \profresult{sd.tt('rebo_neigh') / vd.tt('rebo_neigh') /
  100}{2.5}x.

\subsection{Short-Ranged Contributions}\label{ssec:short-force}

Since \profresult{(o.tt('torsion') + o.tt('frebo')) / (o.tt('torsion') + o.tt('frebo') + o.tt('flj') + o.tt('rebo_neigh'))}{4.7}\
contributions, the optimization of this segment is important
for the overall speedup (it is also crucial for users who want to use REBO instead of AIREBO).

At a first glance, vectorization appears to be impossible, as short neighbor
lists contain few elements (typically $\leq 4$), and 
consequently, the loops over these atoms also have low trip counts.
Indeed, since the code always skips atoms from outer loops, these trip counts
are even lower than the number of elements in the neighbor lists.
It is therefore impossible to achieve good utilization of vector units, if not

by vectorizing along the outermost loop (line 1 in Alg.~\ref{alg:rebo}), which iterates along all local atoms; this approach is
effective since the number of local atoms is much larger than the vector width.

For the short-ranged vectorization, four ideas are important:
1) The ``customary'' loop---i.e., the loop over interaction partners---will not lead to good vector efficiency;
in realistic simulations, this loop is too short, having at most four entries,
and a loop trip count of four is too limiting for all modern computing architectures.

Instead, we vectorize along interactions, i.e., pairs of atoms that are supposed to interact (i.e. lines 1 and 2 in Alg.~\ref{alg:rebo}).
2) These interactions can also be sorted by their constituent atom species.
This allows us to calculate vectors of carbon/carbon, carbon/hydrogen, \dots, interactions individually.
Given that the parameters of the interaction are primarily dependent on the
species, parameters can be read via broadcasts instead of a multi-instruction sequence of blends.
Additionally, the atom species also tend to predict the behaviour in the bond-order calculation, i.e., the neighbor lists of constituent atoms tend to be similar in length.
3) The bond-order calculation is merged into the routine, to maximize the amount of data shared between the different calculation stages.
4) The torsion and REBO force calculation are merged since they operate on the same data.

We achieve a speedup of \profresult{(o.tt('frebo') + o.tt('torsion')) /
  sd.tt('frebo') / 100}{1.0}x from just the optimizations (batching, data reuse and loop merging), i.e. overheads and speedups cancel, and another factor of \profresult{sd.tt('frebo') / vd.tt('frebo')
  / 100}{5.4}x from vectorization.

\begin{figure*}
\centering
\includegraphics[width=\linewidth]{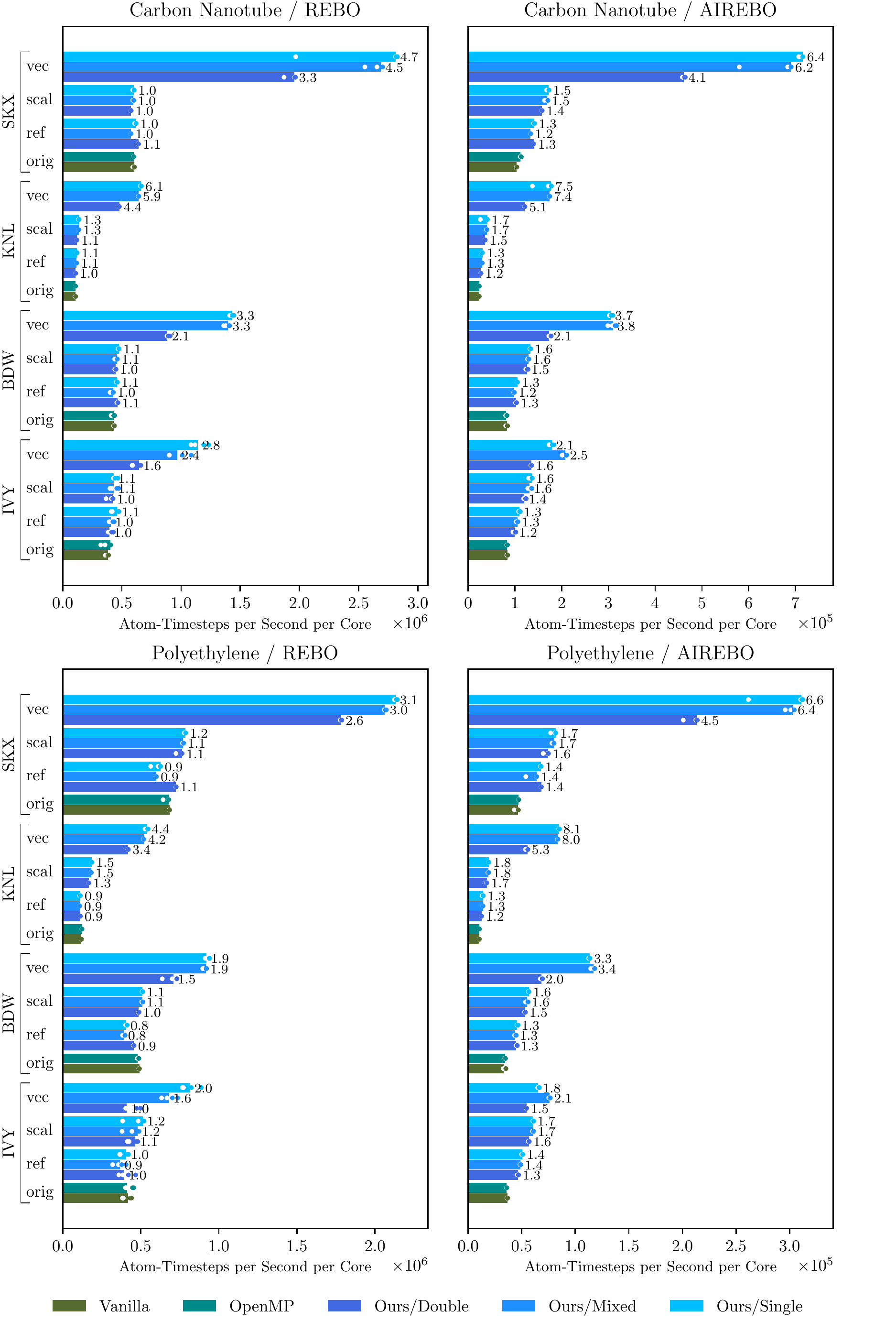}
\vspace*{-0.8cm}
\caption{Performance of the benchmarked systems using REBO and AIREBO with both
  simulations, single thread. Speedups in gray. 
``orig'': original code, ``ref'': our refactored code, ``scal'': our optimized code with scalar backend, ``vec'': our optimized code with vectorized backend.
Performance of the benchmarked systems using AIREBO with both simulations, full node.
For ``Carbon Nanotube'' 5720 atoms, for ``Polyethylene'' 3600 atoms.}
\label{fig:thr-airebo}
\label{fig:thr-rebo}
\end{figure*}

\subsection{Longer-Ranged Contributions}

The vectorization of the inter-molecular forces itself is more straightforward than the short-ranged vectorization.
As the longer-ranged neighbor list includes a skin distance, a cutoff check is
necessary; on architectures that support this feature\footnote{AVX2 supports for
  reasonably fast emulation.},
the vectors are compressed to avoid excessive masking.

The method also needs to handle the two exclusion criteria,
one based on $b^*_{ij}$, and another based on $C_{ij}$.
We start the discussion with the term $b^*_{ij}$, the computation of which is 

only necessary if the two atoms $i$ and $j$ are within a certain range from each other, given by the cutoff function $S(t_r(r_{ij}))$.
Both the original and our code evaluate said function, and calculates $b^*_{ij}$ only if it is non-zero.
It is important to only perform this check after calculating $C_{ij}$, as many of the atoms that are excluded due to $C_{ij} = 0$ are close enough to $i$ that the bond-order would seem necessary.
Since only few interactions actually require the bond-order, calculations that
require it are deferred until sufficiently many such pairs are found, at which
point they are calculated in the same manner as short-ranged forces, taking
advantage of the same tricks as for bond-order calculation.{as described in Sec.~\ref{ssec:short-force}.}

In the original LAMMPS code, the computation of $C_{ij}$ involves a nested loop that
searches for the first connection for which $C_{ij} = 0$, and otherwise for the minimum $C_{ij}$;
if $C_{ij} = 0$, it breaks early.
The code also contains checks to make sure that the calculation is only performed if necessary, i.e., the distance between atom $i$ and atom $j$ is neither too short that a connection exists without any intermediate (in which case $C_{ij} = 0$ always), nor that they are too far apart for such a connection to possibly exist (in which case $C_{ij} = 1$ always).

In all other cases, the code performs the search for a connection between atoms $i$ and atoms $j$ as follows:
It loops first over neighbors of atom $i$ and then over the neighbors of the neighbors of atom $i$.
In each loop, it checks if these neighbors and neighbors of neighbors are connected to the atom $j$.
For example, assuming three neighbors per atom (typical for a carbon
nanotube), the worst case would result in $3 + 3\cdot 3$ checks performed.

Our approach exploits the fact that since the number of short-ranged neighbors
is small, the number of elements that could have non-unity $C_{ij}$ is also small.
Assuming an upper bound of 8 short-ranged neighbors per atom, how may{many} such elements could there be?
Any atom would have 8 direct neighbors, these in turn would have 8 neighbors, and those again 8;
in total, there could be $8 + 8^2 + 8^3 = 584$ neighbors.

Our optimization approach consists in{of} reversing the procedure:
Instead of performing a search for connections between pairs of atoms $i$ and $j$, we map out all the atoms $j$ connected to atom $i$ once, and use that map throughout.
To this end, we use a hashmap (open addressing, 1024 entries) and a buffer of
associated data (needed when $C_{ij}$ is neither one nor zero; as this rarely
happens, it contains just 128 entries); as
usual, a sequential fallback ensures correctness even in unlikely cases.
The search for connections is itself sequential, as it involves insertion into the hashmap.
However, the subsequent lookups are simple to vectorize, a perk that motivated this design.

Since the Lennard-Jones
cutoff is much larger than three times the short-ranged cutoff, all the
elements that are entered into the hashmap will also be checked for it. This
means that in practice this optimization always results in a gain.
Indeed it speeds up the path calculation by \profresult{(o.tt('path_forward') + o.tt('path_reverse')) / (sd.tt('path_forward') + sd.tt('path_search') + sd.tt('path_reverse')) / 100}{3.6} and gains an additional factor of \profresult{(sd.tt('path_search') + sd.tt('path_forward') + sd.tt('path_reverse')) / (vd.tt('path_forward') + vd.tt('path_search') + vd.tt('path_reverse')) / 100}{1.2} from vectorization.
The vectorization speedup is low because it includes the time spent
constructing the hashmap, a purely sequential task.

\subsection{Performance Portability}\label{ssec:port}

From our refactored code, we first created an architecture-specific
implementation that targeted double precision AVX-512.
This choice allowed us to be as close to the hardware as possible,
and since no abstraction hindered our access to the underlying primitives, 
it also aided debuggability. However, once reliable vectorized code is
available, it is desirable to be independent of the intrinsics, because on the one
hand we wanted support also for single precision and mixed precision,
and on the other hand our goal was to be performance-portable on all current x86 CPUs, i.e., AVX, AVX2, IMCI and AVX-512.

The challenge lay{lies} in the sheer size of the code, which made the manual 
translation to a more portable form a task both daunting and extremely
error prone.

As a target for better performance-portability, we chose a custom vector class library that wraps the compiler intrinsics \cite{fog_vcl_2015}.

Instead of a manual translation, we chose to write a simple, clang-based source-to-source rewriting tool.
It replaced calls to{occurrences of} intrinsics with calls to the abstraction library, variable declarations of intrinsic type with vector-class-type variable declarations, and magic numbers derived from the vector length with appropriate expressions.
As a clang tool, it rewrites only specific portions of the code, and leaves
the rest, e.g., formatting and comments, intact.

We can then implement backends for all the desired precisions and architectures.
Compared to manually porting the entire code, the addition of a new backend is
a relatively simple and highly mechanical job.
In particular, we provide a ``scalar'' backend that does not require any compiler support.
This allows us to later evaluate the effect of vectorization against all the other
optimizations, and (optimized, but not vectorized) portability to any architecture.

A backend needs to implement a number of standard mathematical operations, as well as vector gather, scatter, any/all mask checks, compress and expand operations.

The latter three items, mask checks, compress and expand, are crucial for the optimization, and the missing elements that hinder a successful optimization using just standard programming models such as OpenMP.
In the past, conflicting writes were also an issue, but OpenMP 4.5 can perform this with the ``ordered simd'' construct.
A mask check would allow a program to check if a condition holds for any or all active lanes in a OpenMP simd region;
effectively, this is the equivalent of the ``\_\_all''/``\_\_any'' constructs in CUDA.
Current work aims to enable compress and expand patterns with OpenMP, as
it would prove useful to fill up lanes in any loop that contains either ``continue'' or ``break'' statements \cite{krzikalla_dynamic_2016}.

In principle, porting the code to other vector architectures would not be a problem.
However, given the observed AVX speedups, backends for any 128-bit/4-element vector architecture such as NEON (ARM), VSX (POWER) or QPX (BlueGene) promise little pay-off.
On the other hand, the upcoming ARM SVE extension might be
a worthwhile target as it is 2048-bits wide \cite{lee_extending_2017}.
A GPU version, replacing mask checks by the aforementioned CUDA intrinsics, might also lead to sizeable speedups, but would require considerable changes w.r.t.~memory management due to the way neighbor lists are structured in AIREBO.

\section{Results}\label{sec:results}

We assess the performance of our code on two model
problems and four computer architectures, in multiple 
scaling regimes.

Following widely accepted benchmarking principles
\cite{hoefler_scientific_2015}, every experiment is repeated 5 times, and report the median{we repeat every experiment 5 times and report the median}, while plotting the other measurement results as error indicators.
As our performance metric (i.e. higher is better), we choose atom-timesteps per second, because it normalizes both the number of atoms and the timestep.
As such, it is most widely comparable to other systems and circumstances.

We perform the benchmarks on four computing systems, labelled as IVY, BDW, KNL, and
SKX.
\begin{itemize}
\item IVY is a cluster consisting of 27 dual socket Intel Xeon E5-2680v2 nodes (10 cores/socket, 2.8GHz base, launched 2013, AVX, Infiniband QDR interconnect);
these are the oldest processors presented in this study.
\item 
BDW is a cluster of 600 dual socket Intel Xeon E5-2650v4 (12 cores/socket, 2.2GHz base, launched 2016, AVX2, Infiniband EDR interconnect) nodes.

\item 
KNL is a cluster of 4200 Intel Xeon Phi 7250 nodes (68 cores, 1.4GHz base, launched 2016, AVX-512, OPA interconnect).

The nodes run in cache quadrant mode.
\item
  
SKX is a cluster of 1700 dual socket Intel Xeon Platinum 8160 nodes (24 cores/socket, 2.1GHz, launched 2017, AVX-512, OPA interconnect).

\end{itemize}
In all instances, we used the Intel Compiler, and ran under Intel MPI.
The LAMMPS version used was ``23 Oct 2017''.

\begin{figure*}
\centering

\includegraphics[width=\linewidth]{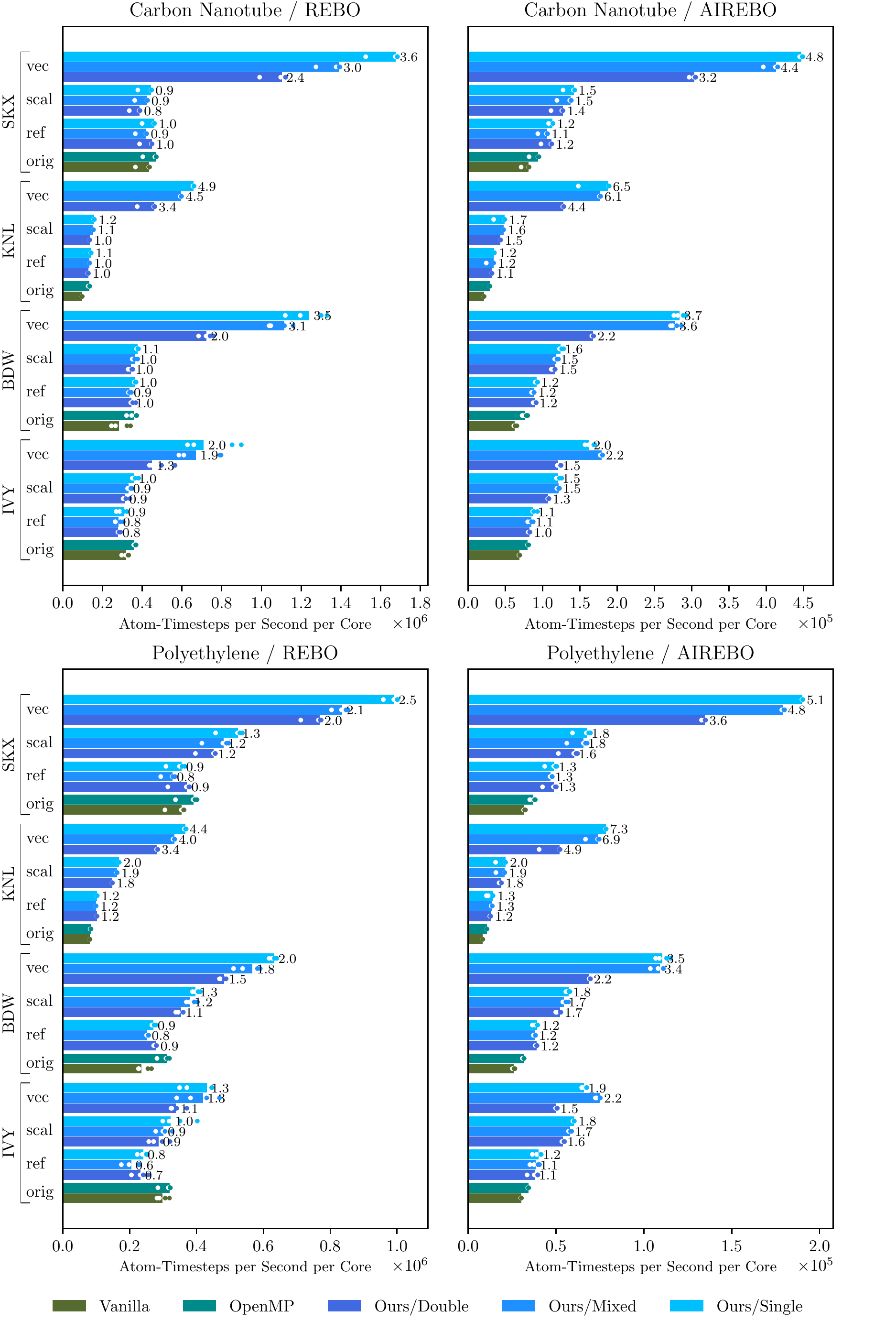}
\vspace*{-0.8cm}
\caption{Performance of the benchmarked systems using REBO and AIREBO with both
  simulations, full node. Speedups in gray. 
``orig'': original code, ``ref'': our refactored code, ``scal'': our optimized code with scalar backend, ``vec'': our optimized code with vectorized backend.
Performance of the benchmarked systems using AIREBO with both simulations, full node.
For ``Carbon Nanotube'' 274560 atoms, for ``Polyethylene'' 172800 atoms.
}
\label{fig:node-airebo}
\label{fig:node-rebo}
\end{figure*}

\subsection{Model Problems}

Due to the complexity of AIREBO, one simulated system might not be
enough to stress all code paths. 

For this reason
we use

the polyethylene benchmark straight from the LAMMPS project, and a custom
carbon nanotube simulation; together, they
exercise different aspects of a simulation:
the first one stressed carbon-carbon, carbon-hydrogen and hydrogen-hydrogen
interactions, while the second contains no hydrogen, and stresses all the dihedral terms in the potential.

We consider four different sizes of simulations, for runs with a single thread or core, a single node, up to 24 nodes, and up to 128 nodes.

\subsection{Accuracy at Lower Precision}

In MD simulations, there are several sources of numerical error,
the most prominent of which is round-off, 
which occurs both in the force calculation and the integration routines.
A discretization error also stems from the integration of forces and velocities to update positions.
In some simulations the approximation of long-ranged
forces is also source of errors (such approximations are mostly used for electrostatics, and not a factor in our simulations).

When switching from double to a lower precision, it is obvious to wonder
whether the code attains an accuracy in agreement with the precision.

A simple test to check the accuracy of an MD simulation is whether or
not it conserves energy.
Any correctly implemented potential should conserve energy while performing a reasonable simulation.

Single precision is always  attractive for vectorization:
A single precision non-vectorized code is expected to be only
  marginally faster than its double precision counterpart; the only gains come from reduced memory usage (and better cache utilization) and cheaper  mathematical functions (sine, cosine, \dots).
  With vectorization, the increased vector width leads to a potential 2x speedup.
However, one has to investigate whether or not reduced precision is numerically
  suitable.

For the energy conservation{} test, we allowed the simulation of the benchmarked systems to run for \num[group-separator={,}]{2000000} timesteps.
The energy conservation test was performed with a 0.25ps timestep to further eliminate integration errors.

Fig.~{fig:energy-conservation} highlights these results.
The energy conservation is in an appropriate range everywhere but in the CNT benchmark with single precision, where energy drift occurs.{
In the polyethylene benchmark, all the different precision modes behave similarly: Their energy oscillates around the reference and never exceeds a relative error of 0.005\
For the Carbon Nanotube benchmark, error oscillates and remain below 0.0005\
We conclude that the energy conservation behaviour is acceptable in all cases but the single precision Carbon Nanotube benchmark.}

\begin{figure*}
\centering
\includegraphics[width=\linewidth]{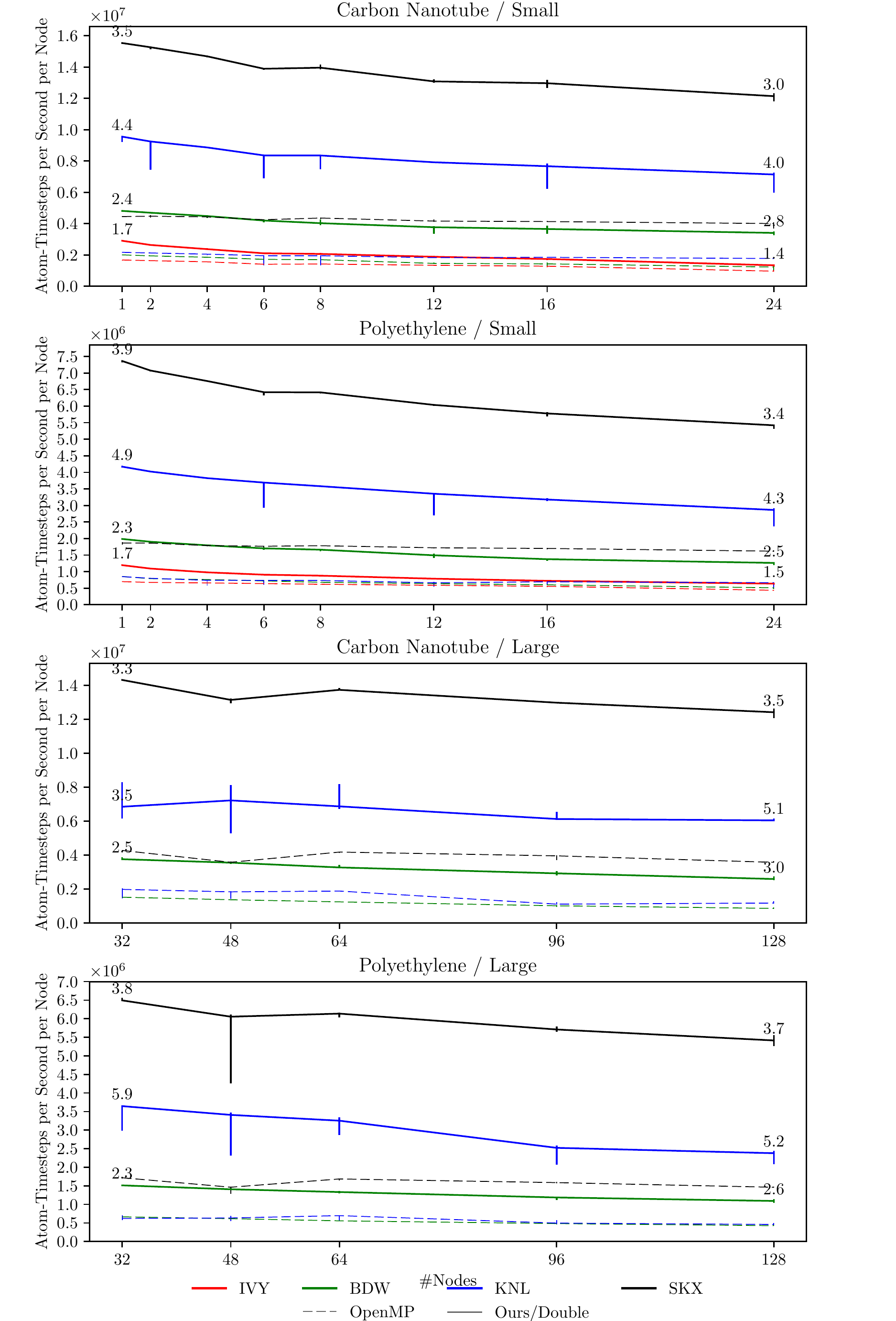}

\vspace*{-0.8cm}
\caption{Strong scaling from 1 to 24 and 32 to 128 nodes with AIREBO and both model problems (for ``Polyethylene'' 1843200 and 14745600 atoms respectively, for ``Carbon Nanotube'' 2928640 and 23429120 atoms respectively). Black numbers are speedups at that scale.}
\label{fig:sc-small}
\label{fig:sc-large}
\end{figure*}

\subsection{Thread-Level Performance}

The first performance test is on a single core with a single thread.
This execution minimizes the overhead due to parallelism and is consequently particularly suitable to estimate the effect of our improvements to single-threaded performance.

Fig.~\ref{fig:thr-airebo} summarizes the performance that our code, in its
three variants (``ref'': refactored code, ``scal'': vectorized code with the scalar backend, ``vec'': vectorized code with the architecture specific vectorized backend), achieves against the reference (``orig''), for the full computation of the AIREBO
potential.
Fig.~\ref{fig:thr-rebo} summarizes the same results, but limited to the REBO part of the potential.
We do not provide separate results for AIREBO-m, since they are nearly identical to

Fig.~\ref{fig:thr-airebo}.

The performance for ``ref'' does not differ a lot
from the previous LAMMPS implementation; this implies that the refactoring had a negligible performance impact.
The performance of ``scal''  shows how our optimizations behave without vectorization;
it is also an indicator of the  vectorization overhead.
In most cases, the improvements due to our optimizations exceed the overheads
by a margin of 5 to 30\

The overall speedup achieved with vectorization depends on the benchmark, potential, and system.
The speedups are increasingly better on the IVY system, then BDW, SKX and finally KNL.
This matches the expectation, since the KNL system is meant to be used with AVX-512.
The speedup tends to be a bit larger in the CNT{Carbon Nanotube} benchmark, but the effect is relatively small.
The overall performance however differs between the two: The CNT{Carbon Nanotube} benchmark is a lot cheaper on a per-atom basis, whether optimized or not.
The speedups for REBO are smaller than for AIREBO since REBO spends a larger fraction of the time doing neighbor list calculations, which do not vectorize well.

\subsection{Core-Level Performance}

Here we present results that take advantage of hyper-threading.
The code is still pinned to a single physical core (and all its logical CPUs), but we utilize multiple threads.
We investigated the full range of possible hyper-thread amounts, and present the performance with the best setting:
for our code, two hyper-threads everywhere; for the OpenMP reference, four hyper-threads on KNL, and two hyper-threads everywhere else.

Hyper-threading increases performance
pretty much across the board.
A typical increase is between 5\
Since the results so closely resemble Fig.~\ref{fig:thr-airebo}
and Fig.~\ref{fig:thr-rebo}, no plot is shown. 

\subsection{Node-Level Performance}

We present scaling result up to one entire node.
As such, these runs utilize as many MPI ranks as there are cores in the architecture.
These tests provide insights into the scaling behaviour, and into parallel overheads while excluding interconnect performance.
For relatively small simulations, this will be the most likely mode of operation, as it is the smallest unit that can be requested in many supercomputing centers.

As apparent from Fig.~\ref{fig:node-airebo} and Fig.~\ref{fig:node-rebo}, the simulation rate is approximately halved and speedups are slightly reduced compared to single-threaded runs.
This result is not surprising, since parallelization means unavoidable
overheads to which our code is particularly exposed since it spends relatively more time communicating.

Nevertheless, the speedups are 1.5x (IVY), 2.2 (BDW), 4.9 (KNL) and 3.6 (SKX) in double precision, and
2.2x, 3.4x, 6.9x and 4.8x in mixed precision.

\subsection{Cluster-Level Performance}

Finally, we present strong scaling results from a single node to up to 24 nodes (Fig.~\ref{fig:sc-small}), and from 32 nodes to 128 nodes (Fig.~\ref{fig:sc-large}, IVY not shown since it is too small).
Due to the considerable resources that such benchmarking runs take, we limit ourselves to the AIREBO potential itself, the fully vectorized code, and double precision.

It appears that both KNL and SKX experience large variance when benchmarked with 48 nodes.
Both of these systems are housed in the same datacenter, and it is possible that this presents an edge-case in terms of networking setup or domain decomposition.

As expected, the performance decreases when scaling up, and the effect is more pronounced in the optimized case since it spends (relatively) less time on computation and more time on communication.
Nevertheless, the speedups 
are still sizeable: more than 2x for BDW, 3x for SKX, and 4x on KNL.

\section{Conclusion}\label{sec:conclusion}

We covered the entire process to achieve high performance on LAMMPS'
  AIREBO, a complex legacy
code, while also carrying out non-trivial debugging of the original code as
well as and code refactoring.
Due to the complexity of the code, software engineering practices are crucial to make this kind
of optimization process feasible.
We presented an innovative approach to make intrinsics-based code more
portable, as well as multiple specific ways to render the target code more vectorizable.

Our optimized code is up to 8x faster, and sustains large-scale, same-precision speed-ups between 3x and 4x.
The used techniques will apply almost 1-to-1 to the optimization of other many-body potentials as well.

As an outlook on future activities, 
we provided insights for both possible OpenMP features and requirements for vector class libraries;

furthermore, we are looking into tools to insure correctness of code for 
many-body potentials by construction.

\section*{Open Source}
Our code is distributed with LAMMPS in an optional module, so that users with supported hardware can enable it.
As part of LAMMPS, the code is licensed under GPLv2 and available on Github.

\section*{Acknowledgements}

Simulations were performed with computing resources granted by RWTH Aachen University under project nova0013.
The authors acknowledge the Texas Advanced Computing Center (TACC) at The University of Texas at Austin for providing HPC resources that have contributed to the research results reported within this paper.
We are grateful to
Cyril Falvo, Efrem Braun, Marvin Kammler, and the LAMMPS authors
for help in 
the bug-hunting.

\bibliographystyle{siamplain}

\bibliography{biblio}

\end{document}